\documentclass[a4paper,
               ]{jacow}
\usepackage{pdfpages,multirow,ragged2e} %
\usepackage{enumerate}
\usepackage{caption}
\usepackage{subcaption}
\makeatletter%
	\ifboolexpr{bool{xetex}}
	 {\renewcommand{\Gin@extensions}{.pdf,%
	                    .png,.jpg,.bmp,.pict,.tif,.psd,.mac,.sga,.tga,.gif,%
	                    .eps,.ps,%
	                    }}{}
\makeatother

\ifboolexpr{bool{xetex} or bool{luatex}} % test for XeTeX/LuaTeX
 {}                                      % input encoding is utf8 by default
 {\usepackage[utf8]{inputenc}}           % switch to utf8

\usepackage[USenglish]{babel}

\ifboolexpr{bool{jacowbiblatex}}%
 {%
  \addbibresource{jacow-test.bib}
  \addbibresource{biblatex-examples.bib}
 }{}
\begin{document}

\title{Quench Detection in a Superconducting Radio Frequency Cavity with Combined Temperature and Magnetic Field Mapping\thanks{The work is partially supported by the U.S.Department of Energy, Office of Science, Office of High Energy Physics under Awards No. DE-SC 0009960, NSF Grant 100614-010, and  by the U.S. Department of Energy, Office of Science, Office of Nuclear Physics under contract DE-AC05-06OR23177.}}

\author{B. D. Khanal\textsuperscript{1}\thanks{bkhan001@odu.edu},   P. Dhakal\textsuperscript{2}, and G. Ciovati\textsuperscript{1,2} \\
				\textsuperscript{1}Center for Accelerator Science, Department of Physics, \\Old Dominion University, Norfolk, VA 23529, USA\\
                \textsuperscript{2}Thomas Jefferson National Accelerator Facility, Newport News, VA 23606, USA \\ 
		}
	
\maketitle

\begin{abstract}
Local dissipation of RF power in superconducting radio frequency cavities create so called "hot-spots", primary precursors of cavity quench driven by either thermal or magnetic instability. These hot-spots are detected by a temperature mapping system, and a large increase in temperature on the outer surface is detected during cavity quench events. Here, we have used combined magnetic and temperature mapping systems using anisotropic magneto-resistance (AMR) sensors and carbon resisters to locate the hot spots and areas with high trapped flux on a 3.0 GHz single-cell Nb cavity during the RF tests at 2.0 K. The quench location and hot spots were detected near the equator when the residual magnetic field in the Dewar is kept < 1 mG. The hot spots and quench locations moved when the magnetic field is trapped locally, as detected by T-mapping system. No significant dynamics of trapped flux is detected by AMR sensors, however change in magnetic flux during cavity quench is detected by a flux gate magnetometer, close to the quench location. The result provide the direct evidence of hot spots and quench events due to localized trapped vortices. 
\end{abstract}

\section{Introduction}
Superconducting radio frequency (SRF) cavities made from elemental niobium are the building blocks of modern particle accelerators, superconducting electronics and quantum computers because of their formability in complex structures and lithographic thin films. Recent advances in surface engineering by doping and annealing led to an unprecedented quality factor \cite{dhakal13, anna, dhakalreview}, however the process is vulnerable to residual flux trapping when the cavity transitions to superconducting state during cooldown \cite{dhakal20}. The trapped vortices within the RF penetration depth oscillates under the RF field and lowers the quality factor with additional RF dissipation.

The temperature mapping technique is able to detect the regions of large RF power dissipation, referred as "hot-spots" \cite{knobloch}. The source of hot-spots could be due to the normal conducting inclusion or segregation of impurities in dislocations sites or grain boundaries. If the origin of hot spot is due to trapped vortices, those could be moved or change their intensity an application of a thermal gradient \cite{gigi08, gurevichgigi}. The combined temperature and magnetic field mapping system used in 1.3 GHz cavity showed an enhancement of magnetic field trapping due to quench events either by trapping of thermo current induced flux or trapping to additional residual flux \cite{ishwarisrf23, kek22} . Here, we present the results of RF measurements done on a 3.0 GHz single cell cavity with combined temperature and magnetic field mapping where the hot-spots are created on the cavity surface by trapping magnetic field locally.

\section{Experimental setup}
The 3 GHz single cell cavity used in this study was made from high purity large grain Nb and scaled down from the TESLA end-cell cavity shape \cite{aune, gigiprab18}. After previous measurements reported in Ref.\cite{ishwariipac22} the cavity was subjected to $\sim$ 25 $\mu$m electropolishing followed by high pressure rinse with deionized water. The cavity was assembled with input and pick up antennas. A cavity vacuum of < 10$^{-8}$ mbar was maintained with active pumping during the cooldown and RF tests. 

\begin{figure}[!h]
   \centering
   \includegraphics*[width=1\columnwidth]{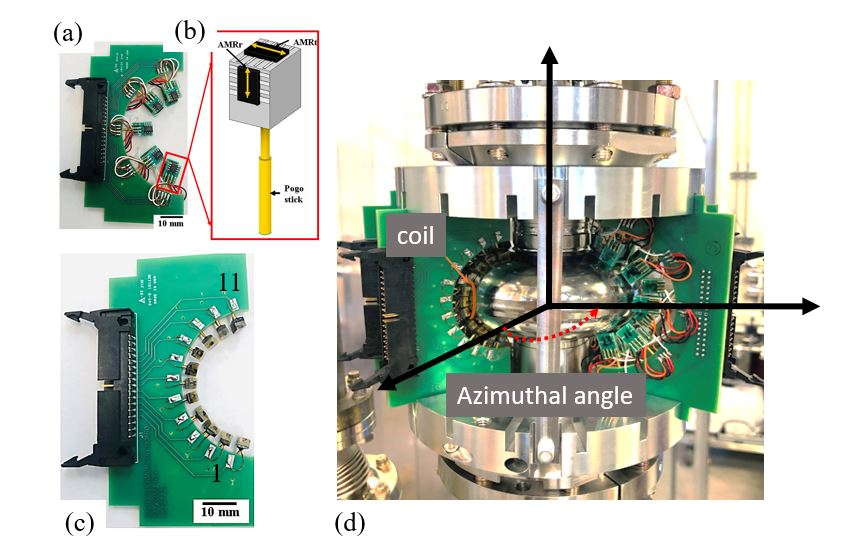}
   \caption{(a) and (b) AMR sensors and (c) temperature sensors board, (d) assembled on the surface of cavity \cite{ishwariphd}.}
   \label{fig:setup}
\end{figure}
The combined temperature and magnetic field mapping system relies on 100 $\Omega$ Allen-Bradely carbon resistors to measure the temperature and AMR sensors to measure the magnetic field. The details of set up, sensors calibration, sensitivity and measurement can be found in Refs. \cite{ishwarisrf, ishwarirsi,ishwariphd}. The AMR sensors boards and temperature sensors boards are place alternatively on every 22.5$^\circ$ around the cavity. There are 10 AMR sensors on each board measure the radial and tangential component of the magnetic field in the vicinity of cavity's outer surface. 11 temperature sensors are mounted on each board which are kept in contact with the cavity surface with spring-loaded Pogo sticks. Apiezon-N cryogenic vacuum grease was applied on temperature sensors to make good thermal contact with the cavity surface. Magnetic sensor \#3 and temperature sensors \# 6 lies on the equator of the cavity. Three flux gate magnetometers (FGM) and 4 Cernox sensors were attached on the cavity surface to measure the residual magnetic field and temperature of outer cavity surface and helium bath. One of FGM was place away from the cavity surface and parallel to the cavity axis to measure the residual magnetic field on the Dewar. Two FGM were placed along the axis of small coils and perpendicular to the cavity axis  as shown in Fig. \ref{fig:setup}. The Cernox sensor were place on the two beam tube, closer to the cavity irises. The carbon resistors were calibrated against the Cernox sensors for each measurements. 

The cryogenic Dewar is equipped with compensation coils capable of maintaining the residual magnetic field on cavity under test < 1 mG. In our current set up, two small coils  of radius $\sim$ 10 mm with 10 turns made from 38 AWG enameled copper wire were mounted on the outer surface of the cavity in equatorial region. The coils were mounted in such a way that temperature and magnetic field sensors are placed on the axis of the coil. The magnetic field produced by the coil will be trapped locally when the cavity undergoes superconducting transitions. 

The measurement procedure is as follow: (i) The magnetic field in Dewar was set below 1 mG and offset voltage of AMR sensors was recorded at 10 K. (ii) The cavity was cooled to 4.2 K and filled with liquid helium by maintaining temperature difference of > 2 K between the cavity irises. (iii) $Q_0$(T)  at low rf field (peak surface rf magnetic field $B_p \sim$ 15 mT) from 4.2-2.0 K was measured while calibrating the carbon resistors against the calibrated Cernox sensors.  (iv) $Q_0$ vs $E_{acc}$ at 2.0 K was measured along with the temperature and magnetic field maps. (v) The cavity is warmed-up above $T_c$ ($\sim$ 9.2 K) and 200 mA current on the coil that was mounted on the cavity surface, producing $\sim 100$ $\mu$T magnetic field perpendicular to the cavity surface as measured by FGM. (vi) The cavity was cooled down with less than 0.1 K temperature difference between the cavity irises in order to maximize the flux trapping. When the cavity reached superconducting state ($\sim$ 4.2 K), the current to the coiled is switched off. The magnetic field reading after the current being turned off is the amount of magnetic field trapped on the cavity surface. It was found that about 40 \% of the magnetic flux is trapped on the cavity surface. 
(vii) Steps (iii) and (iv) were repeated by placing the coil at different locations on cavity surface.

\section{Experimental Results}
\subsection{rf results}
\begin{figure}[!h]
   \centering
   \includegraphics*[width=1\columnwidth]{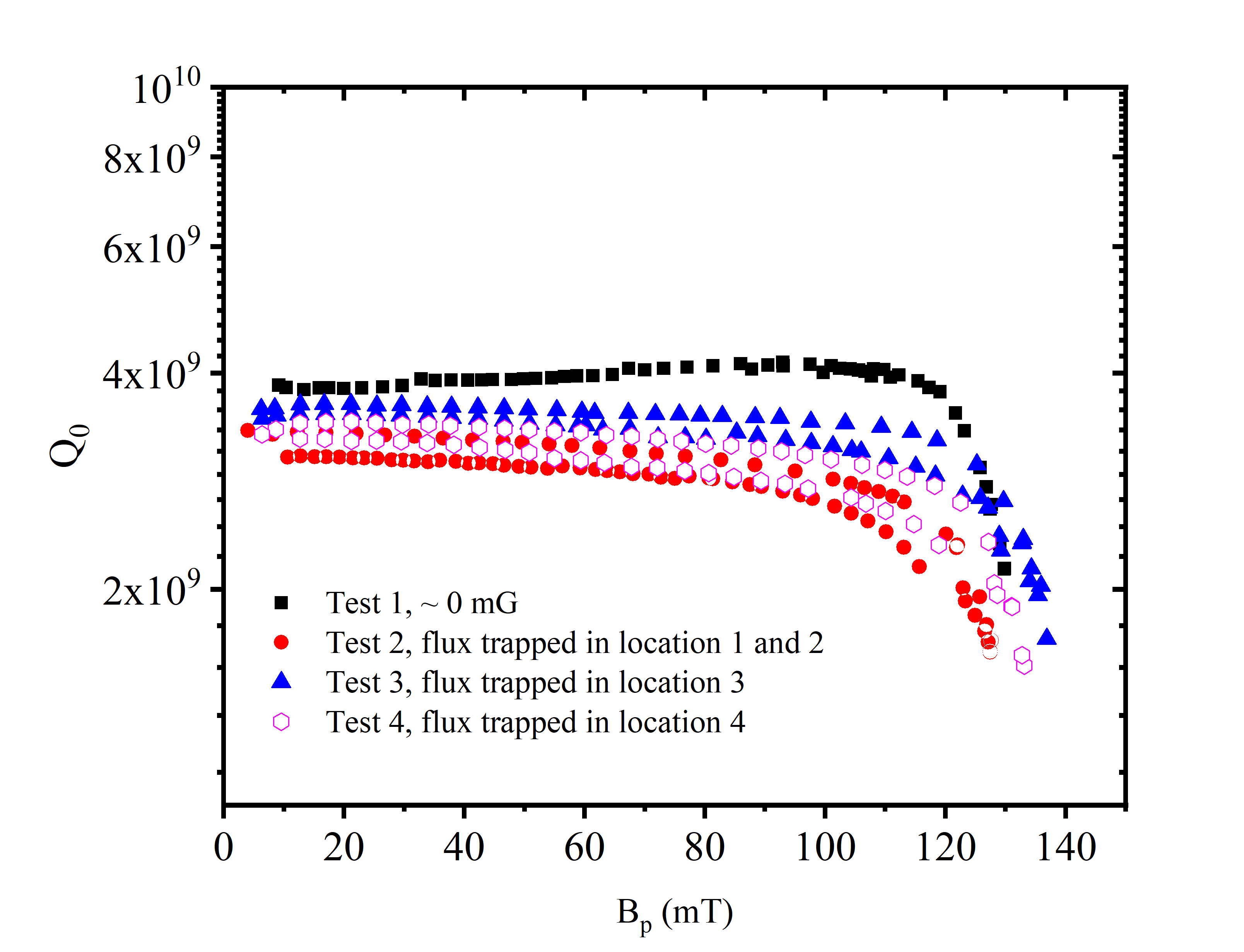}
   \caption{$Q_0(B_p)$ measured at 2 K for different condition of trapped magnetic field. The experimental uncertainties on $Q_0$ and $B_p$ are $<10 \%$, and $<5 \%$ respectively. All rf tests were limited by quench.}
   \label{fig:QBp}
\end{figure}

Figure \ref{fig:QBp} shows the $Q_0(B_p)$ measured for 4 different flux trapping condition. The test labeled test 1 is after the first cooldown with < 1 mG residual flux during the cooldown. The low field $Q_0$ is $\sim$ 4$\times 10^9$ and increases with the increase in RF field. The Q-rise is similar to the doped niobium cavities \cite{dhakalreview}, however similar results were also reported in electropolished cavities resonating at higher frequencies \cite{martina}. The cavity showed high field Q-slope starting at $B_p \sim$ 110 mT before quenching at $B_p \sim$ 139 mT.

The second RF test, test 2, was done with residual flux being trapped on two different locations. Location 1 corresponds to the 0$^\circ$ (reference point for B-T mapping system), whereas location 2 corresponds to 202.5$^\circ$. The quality factor was lowered by $\sim$25 \% and  $Q_0(B_p)$ doesn't show any Q-rise phenomenon. The cavity encounter a sudden quench at $B_p \sim$ 116 mT and quality factor jumped to higher value and cavity finally quenched at $B_p \sim$ 127 mT. The measurement was repeated by moving the coil to location 3 (67. 5$^\circ$) and location 4 (202.5$^\circ$). In both cases, the cavity quenched at $B_p \sim$ 135 mT with hysteresis in $Q_0(B_p)$, due to the release of trapped flux during quench. The flux gate magnetometer reads a drop in trapped flux by $\sim$ 50 \% after the quench. The drop in flux is due to redistribution of vortices at quench location.

\begin{figure*}[htb]
   \centering
   \includegraphics*[width=2\columnwidth]{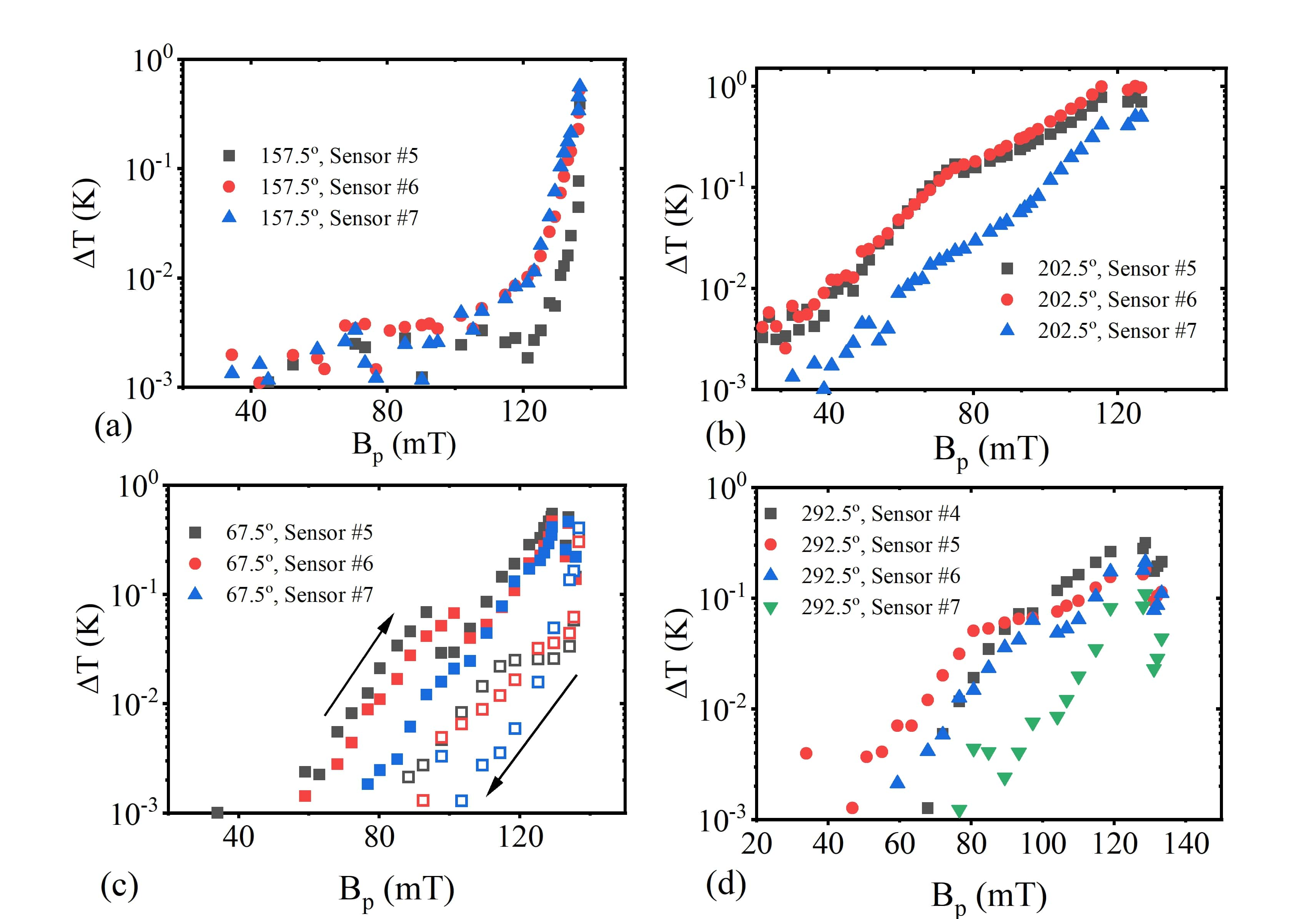}
   \caption{Change in temperature on cavity surface during RF test at 2.0 K measured by the temperature sensors on hot-spots for (a) test 1, (b) test 2, (c) test 3 and (d) test 4. The arrow in (c) represents the temperature during RF power increasing (solid symbols) and decreasing (open symbols). After the quench the hot-spot strength reduces and hysteresis in $\Delta T$ vs $B_p$ is observed. }
   \label{fig:dtvsbp}
\end{figure*}

\subsection{B and T mapping}
\begin{figure}[htb]
   \centering
   \includegraphics*[width=0.9\columnwidth]{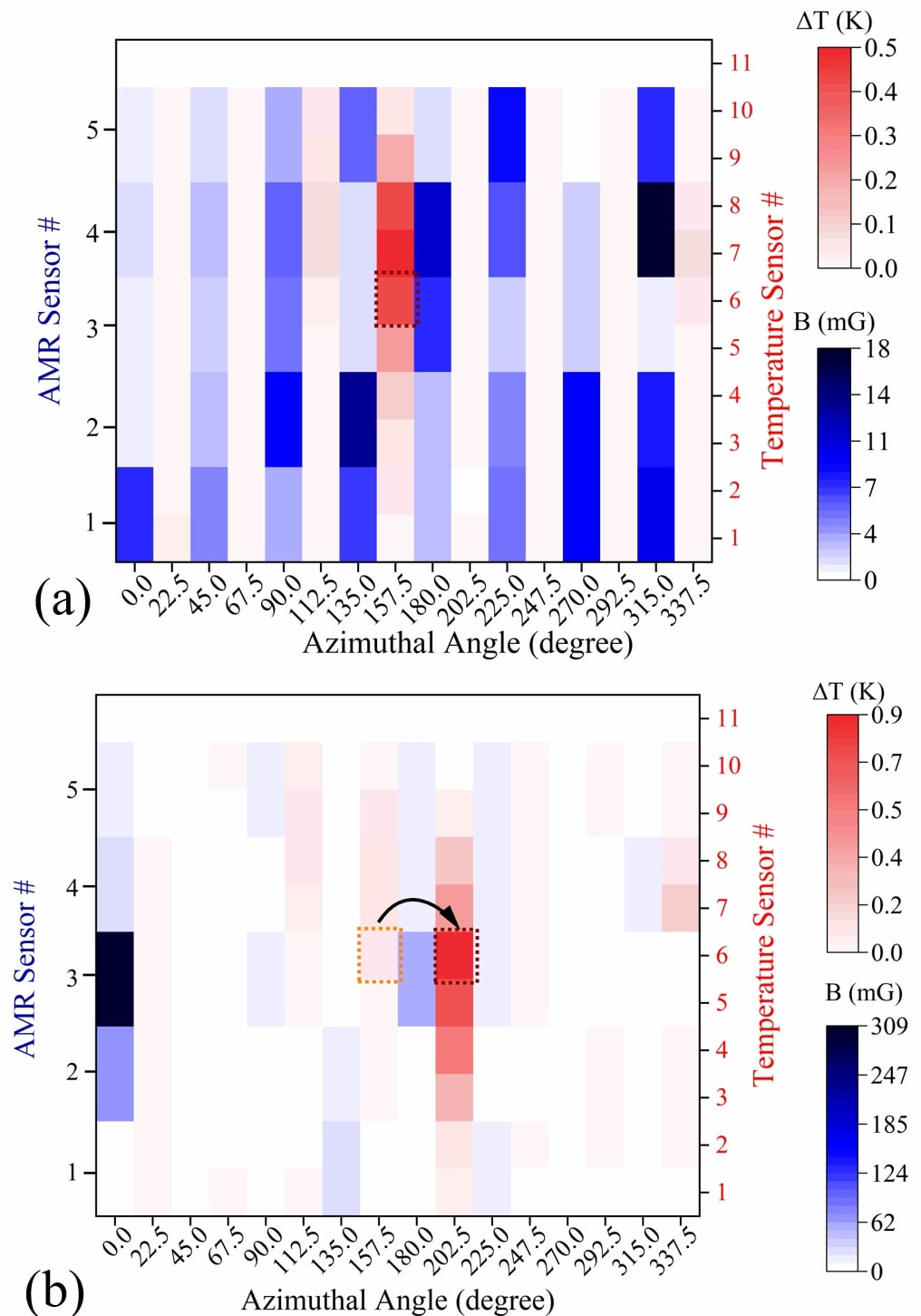}
   \caption{Unfolded temperature and magnetic field  map during the cavity quench for (a) test \#1, and (b) test \#2.  }
   \label{fig:2D}
\end{figure}

The first RF test was done with <1 mG residual magnetic field in the Dewar, measured by flux gate magnetometer and also verified by AMR sensors at 10 K. When the Dewar is filled with liquid helium, a small change in residual field was measured by flux gate and AMR sensors. This change in field doesn't affect the RF performance, since the cavity is already in Meissner state. During the RF test \#1, as shown in Fig. \ref{fig:QBp}, the temperature sensors didn't show any significant change as RF field is increased up to $B_p \sim $ 120 mT. On further increase in RF field, the temperature on sensors \# 5, 6, and 7 at 157.5$^\circ$ showed sudden increase in temperature and the cavity quenched at  $B_p \sim $ 139 mT as shown in Fig.\ref{fig:dtvsbp} (a). The quench location was identified to be at 157.5$^\circ$, sensor \#6 as shown in Fig. \ref{fig:2D}(a). The precursor heating was also measured by the adjacent sensors \# 5 and \#7 as shown in Fig. \ref{fig:dtvsbp}(a). The sensor \#6 is located at the equator of the cavity whereas, senors \#5 and \#7 are located at $\sim$ 5 mm away from the equator weld. The sudden increase in temperature $B_p \sim 120~mT$ on hot-spot locations correspond to the onset of Q-slope in $Q_0(B_p)$ during rf test \#1 as shown in Fig. \ref{fig:QBp}. 

During test \#2, 200 mA current was applied to the coil encircling AMR sensors \#3 at 0$^\circ$ and temperature sensors \# 5, 6, and 7 at  202.5$^\circ$. The total magnetic field $\sim$ 300 mG was measured during the RF test, which corresponds to the trapped magnetic field on AMR sensor \#3 at location 0$^\circ$. On increasing RF power during test \#2, no significant change in magnetic field was detected by the AMR sensors at 0$^\circ$, however, temperature sensors at 202.5$^\circ$ detected an increase in temperature starting $B_p \sim $ 50 mT. On further increase in RF field, the temperature increased until the cavity quench first at $B_p \sim$ 117 mT and repetitive quench at $B_p \sim$ 127 mT. Based on the T-mapping data, the quench location is identified to be sensor \#6 at 202.5$^\circ$ as shown in Fig. \ref{fig:2D}, shifted from sensor \#6 at 157.5$^\circ$ as indicated by arrow.  It is unknown that if there is any quench events occurred at 0$^\circ$ due to no change in magnetic field measured by AMR sensors. This result showed the clear demonstration of hot-spot and cavity quench due to locally trapped magnetic field. 

To further investigate the effect of local trapped flux on the hot-spots and quench, we placed the coils at azimuthal angles 67.5$^\circ$ and 292.5$^\circ$ encircling temperature sensors \# 5, 6, and 7. Again, the rise in temperature was measured at the particular location corresponding to the magnetic field trapped as shown in Fig. \ref{fig:dtvsbp}. Interestingly, when the cavity quenched, some trapped flux is released or migrated away from the sensors, but not far enough to be detected by AMR sensors on boards neighboring the T-sensors board.

\section{Discussion and Summary}
\begin{figure}[htb]
   \centering
   \includegraphics*[width=1\columnwidth]{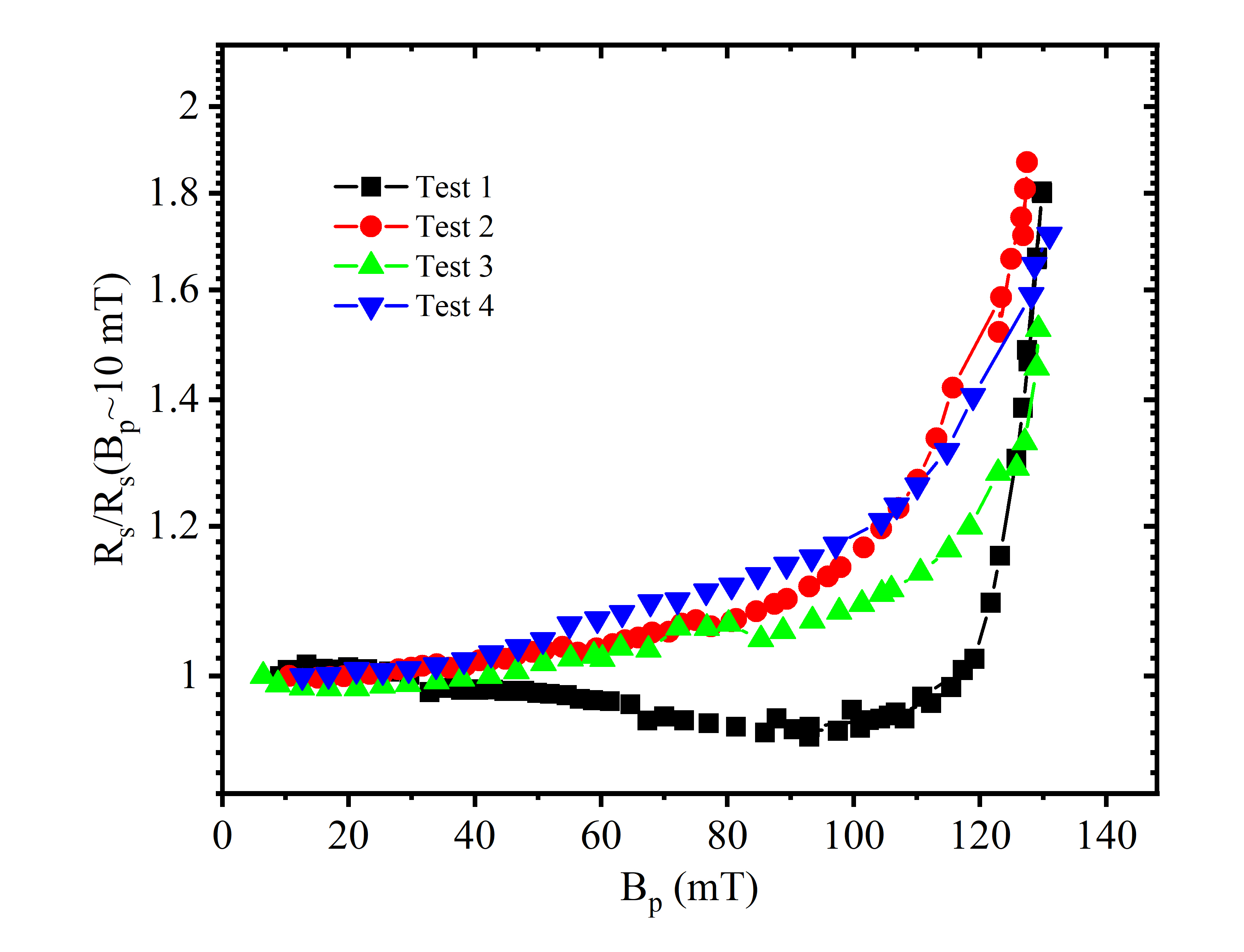}
   \caption{Normalized surface resistance as a function of peak RF magnetic field. }
   \label{fig:RsBp}
\end{figure}

Figure \ref{fig:RsBp} shows the normalized surface resistance as a function of the peak RF field. The surface resistance decrease with the increase in RF field during test 1, typically observed in impurity doped cavities, even though no doping treatment was applied to the cavity. However with the trapped flux, the surface resistance increases with the increase in RF field. The field at which the Q-slope starts corresponds to the onset of temperature increase as shown in Figs. \ref{fig:QBp} and \ref{fig:dtvsbp}(a), which demonstrate that the medium field Q-slope is a result of addition RF loss due to trapped vortices.  The RF loss in hot-spots can be characterized as $\Delta T \sim B_p^n$. The $\Delta T$ vs. $B_p$ was fitted with least square fits. The exponent is found to be n > 4 in the case of local trapped flux suggesting the RF loss due to vortices is stronger than the ohmic-type (n =2) loss. 

 The combined temperature and magnetic field mapping system allows us to identify the hot-spots and quench location on 3.0 GHz cavity. Even though AMR sensors didn't show any significant change in magnetic flux distribution, the flux gate magnetometer mounted near the trapped flux location showed the re-distribution or release of vortices as a result of cavity quench. After the cavity quenched first time at the trapped flux location, the strength of the hot-spots decreases as seen from the drop in temperature of sensors as shown in Fig. \ref{fig:dtvsbp} (c). Further studies are planned to investigate the vortex induced hot-spots and quench locations in 1.3 GHz and 3.0 GHz cavities subjected to different surface preparations. 

\section{ACKNOWLEDGEMENTS}
 We would like to acknowledge Peter Owen for RF and Justin Kent for cryogenic support as well as technical staff members for the cavity fabrication, processing and assembly.

%
% only for "biblatex"
%
\ifboolexpr{bool{jacowbiblatex}}%
	{\printbibliography}%

\begin{thebibliography}{99} % Use for 1-9 references
\bibitem{dhakal13}
P. Dhakal \textit{et al}., "Effect of high temperature heat treatments on the quality factor of a large-grain superconducting radio-frequency niobium cavity", \textit{Phys. Rev. ST Accel. Beams} \textbf{16}, 042001 (2013).
\bibitem{anna}
A. Grassellino \textit{et al.}, “Unprecedented quality factors at accelerating gradients up to 45 MV$m^{-1}$ in niobium superconducting resonators via low temperature nitrogen infusion,” \textit{Supercond. Sci. Technol.}, \textbf{30}, 094004 (2017).
\bibitem{dhakalreview}
P. Dhakal, "Nitrogen doping and infusion in SRF cavities: A review", \textit{Physics Open}, \textbf{5}, 100034 (2020).
 \bibitem{dhakal20}
 P. Dhakal, G. Ciovati, and A. Gurevich. “Flux expulsion in niobium superconducting radio-frequency cavities of different purity and essential contributions to the flux sensitivity”, \textit{Phys. Rev. Accel. Beams,} \textbf{23} 023102 (2020).
 \bibitem{knobloch}
 J. Knobloch, H. Muller, and H. Padamsee, “Design of a high speed, high resolution thermometry system for 1.5 GHz superconducting radio frequency cavities,” \textit{Rev. Sci. Instrum.} \textbf{65}, 3521–3527 (1994).
 \bibitem{gigi08}
 G. Ciovati, and A. Gurevich, "Evidence of high-field radio-frequency hot spots due to trapped vortices in niobium cavities", \textit{Phys. Rev. ST Accel. Beams} \textbf{11}, 122001 (2008).
  \bibitem{gurevichgigi}
 A. Gurevich, and G. Ciovati, " Effect of vortex hotspots on the radio-frequency surface resistance of superconductors", \textit{ Phys. Rev. B} \textbf{87}, 054502 (2013).
 \bibitem{ishwarisrf23}
 I. Parajuli, G. Ciovati, J. Delayen, and A. Gurevich, "Magnetic field mapping of a large-grain 1.3 GHz single cell cavity", this conference, paper ID MOPMB036.
 \bibitem{kek22}
 T. Okada, E. Kako, M. Masuzawa, H. Sakai, R. Ueki, K. Umemori, and T. Tajima, "Observation of quenching-induced magnetic flux trapping using a magnetic field and temperature mapping system", 
 \textit{Phys. Rev. Accel. Beams} \textbf{25}, 082002 (2022).
\bibitem{aune}
B. Aune \textit{et al.}, “Superconducting TESLA cavities”, Physical Review special topics accelerators and beams 3, 092001 (2000).
\bibitem{gigiprab18}
G. Ciovati, G. Eremeev, and F. Hannon, “High field Q slope and the effect of lowtemperature baking at 3 GHz”, Physical Review Accelerators and Beams 21, 012002 (2018).
\bibitem{ishwariipac22}
I.P. Parajuli, G. Ciovati, J.R. Delayen, A.V. Gurevich, and B.D. Khanal, “Preliminary Results of a Magnetic and Temperature Map System for 3 GHz Superconducting Radio Frequency Cavities”, \textit{in Proc. IPAC'22}, Bangkok, Thailand, Jun. 2022, pp. 1315-1318. \url{doi:10.18429/JACoW-IPAC2022-TUPOTK044}.
 \bibitem{ishwariphd}
 I. P. Parajuli, "Characterization of losses in superconducting radio-frequency cavities by combined temperature and magnetic field mapping" PhD Thesis, Department of Physics, Old Dominion University (2022).
\bibitem{ishwarisrf}
I. Parajuli, G. Ciovati, W. Clemens, J. Delayen, A. Gurevich, and J. Nice, “Design and Commissioning of a Magnetic Field Scanning System for SRF Cavities”, in \textit{Proc. SRF’19 , International Conference on RF Superconductivity 19}, (Aug. 2019), pp. 547–549, \url{doi.org/10.18429/JACoW-SRF2019-TUP052}.
\bibitem{ishwarirsi}
I. P. Parajuli, G. Ciovati and J. Delayen, "High resolution diagnostic tools for superconducting radio frequency cavities" \textit{Rev. Sci. Instrum.} \textbf{93}, 113305 (2022).
 \bibitem{martina}
 M. Martinello, M. Checchin, A. Romanenko, A. Grassellino, S. Aderhold, S.K. Chandrasekeran, O. Melnychuk, S. Posen, and D.A. Sergatskov, "Field-Enhanced Superconductivity in High-Frequency Niobium Accelerating Cavities", \textit{Phys. Rev. Lett.} \textbf{121}, 224801 (2018).

 
	\end{thebibliography}
	{
	% "biblatex" is not used, go the "manual" way
	
	%\begin{thebibliography}{99}   % Use for  10-99  references

% end \ifboolexpr
%
% for use as JACoW template the inclusion of the ANNEX parts have been commented out
% to generate the complete documentation please remove the "%" of the next two commands
% 
%\newpage
}

\end{document}